# Multiple Time Step Integrators in *Ab Initio* Molecular Dynamics

Nathan Luehr,[†,‡] Thomas E. Markland[‡] and Todd J. Martínez[†,‡]

[‡]Department of Chemistry and [†]The PULSE Institute
Stanford University, Stanford, CA 94305

**Abstract:** Multiple time-scale algorithms exploit the natural separation of time-scales in chemical systems to greatly accelerate the efficiency of molecular dynamics simulations. Although the utility of these methods in systems where the interactions are described by empirical potentials is now well established, their application to *ab initio* molecular dynamics calculations has been limited by difficulties associated with splitting the *ab initio* potential into fast and slowly varying components. Here we show that such a time-scale separation is possible using two different schemes: one based on fragment decomposition and the other on range separation of the Coulomb operator in the electronic Hamiltonian. We demonstrate for both water clusters and a solvated hydroxide ion that multiple time-scale molecular dynamics allows for outer time steps of 2.5 fs, which are as large as those obtained when such schemes are applied to empirical potentials, while still allowing for bonds to be broken and reformed throughout the dynamics. This permits computational speedups of up to 4.4x, compared to standard Born-Oppenheimer *ab initio* molecular dynamics with a 0.5 fs time step, while maintaining the same energy conservation and accuracy.

**Introduction**

Multiple time step (MTS) integration techniques[1-7] are standard tools used to increase the computational efficiency of molecular dynamics calculations based on empirical force fields. These MTS schemes exploit the fact that the forces in chemical systems can typically be split into fast-varying and slow-varying parts. This splitting is then leveraged to integrate the slow-varying parts with a longer time step and the fast-varying parts with a shorter time step. Since the "slow forces", such as the long range electrostatic interactions, are typically more computationally expensive to evaluate than the "fast forces", such as covalent bond stretches, the ability to evaluate them less often affords significant speed-ups. For empirical potentials this separation is often straightforward as the Hamiltonian is typically written as a sum of terms such as van der Waals, bond stretching, torsional, and electrostatic interactions which can be easily assigned to the "slow" or "fast" part.

In contrast, *ab initio* molecular dynamics (AIMD) schemes compute the potential energy surface on which the nuclei evolve by solving the electronic Schrodinger equation at each time step. This introduces significant flexibility, allowing for bond rearrangement (difficult with empirical force fields that generally assume a prescribed bonding topology),[8] electron transfer,[9] and transitions between electronic states.[10,11] The most straightforward AIMD approach is the Born-Oppenheimer scheme (BOMD).[12-14] In BOMD, the electronic degrees of freedom are assumed to relax adiabatically at each nuclear geometry, and an electronic structure problem is solved fully self-consistently at each time step. An advantage of this approach is that dynamics always occur on a Born-Oppenheimer potential energy surface. In contrast, other AIMD methods employ an extended Lagrangian scheme (Car-Parrinello or CPMD),[15,16] where new fictitious degrees of freedom corresponding to the coefficients in the electronic wavefunction are integrated simultaneously with the nuclear motion. The CPMD method avoids iteration to self-consistency in the solution of the electronic wavefunction at each time step, at the expense of introducing electronic time-scales that are faster than those of atomic motion and thus necessitating smaller time steps for CPMD than would be possible using BOMD. MTS schemes have been developed to mitigate the computational cost of integrating fast electronic degrees of freedom in CPMD.[17-19] These exploit the time-scale separation between the fictitious electronic and nuclear degrees of freedom and allow the outer (nuclear) time step to approach the BOMD limit.

Applying MTS schemes to decompose nuclear motions in AIMD methods presents a much greater challenge since *ab initio* forces do not naturally separate into fast-varying and slow-varying components. Thus, for a long time it has appeared that MTS schemes could not be applied straightforwardly to BOMD (or, equivalently, to the nuclear degrees of freedom in CPMD). Recent work has shown that this conclusion is too harsh, demonstrating that one can treat different components of the electronic structure problem (specifically, the Hartree-Fock and Moller-Plesset contributions) with different time steps in an MTS scheme.[20] This approach leverages the well-known fact that the dynamic electron correlation correction to the Hartree-Fock potential energy surface varies slowly with molecular geometry.

In this paper, we demonstrate two ways of splitting the electronic Hamiltonian that enable AIMD calculations to exploit MTS integrators. The first of these relies on a fragment decomposition of the Hamiltonian. Such fragment decompositions have been previously proposed to accelerate electronic structure computations for large molecular systems.[21-30] In those cases, the energy expression in terms of fragments is viewed as an approximation to the true potential energy surface, and neglected interactions (e.g., relating to charge transfer between fragments) are rarely quantified or corrected. In contrast, our scheme uses the fragment decomposition only as an intermediary representation during inner time steps with corrections included at outer-steps. As such, the dynamics occurs on the potential surface without any fragment approximations.

The second MTS scheme we introduce exploits a splitting of the Coulomb operator in the electronic Hamiltonian. This is closely related to the Coulomb-Attenuated Schrödinger Equation (CASE) approximation that has been proposed to accelerate electronic structure calculations.[31,32] Again, the advantage of our approach is that unlike the CASE approximation, which entirely neglects long-range electrostatic effects, our scheme yields results which are equivalent to those obtained in a calculation employing the usual Coulomb operator while simultaneously allowing for the considerable computational speed-ups afforded by the CASE approach.

**Theory**

Trotter factorization of the Liouville operator provides a systematic approach to derive symplectic time-reversible molecular dynamics integrators for systems containing many time-scales.[3] We begin by briefly reviewing this formalism in order to highlight its desirable properties for the AIMD force splitting schemes presented below. The classical Liouville operator for $n$ degrees of freedom with coordinates, $x_k$, and conjugate momenta, $p_k$ is

$$iL = \sum_{k}^{n}\left[\dot{x}_k \frac{\partial}{\partial x_k} + f_k \frac{\partial}{\partial p_k}\right] \quad (1)$$

where $f_k$ is the force acting on the kth degree of freedom. The classical propagator, $e^{iLT}$, exactly evolves the system by a time period $T$ from an initial phase space point $\Gamma(t) = \{\mathbf{x}(t), \mathbf{p}(t)\}$ at time $t$ to its destination at time $t + T$ via the operation $\Gamma(t+T) = e^{iLT}\Gamma(t)$.

For a multidimensional system with a general choice of interactions, this operation cannot be performed analytically for the full propagator. The Trotter factorization method solves this problem by splitting the Liouville operator,[3]

$$iL = iL_x + iL_p = \sum_{k}^{n}\dot{x}_k \frac{\partial}{\partial x_k} + \sum_{k}^{n} f_k \frac{\partial}{\partial p_k} \quad (2)$$

and applying the symmetric Suzuki-Trotter formula.[33,34]

$$e^{a+b} \approx \left(e^{a/2M} e^{b/M} e^{a/2M}\right)^M + O\left(\frac{1}{M^3}\right) \quad (3)$$

Thus one obtains

$$\Gamma(t+\delta T) = e^{iL_p \delta T/2} e^{iL_x \delta T} e^{iL_p \delta T/2} \Gamma(t) \quad (4)$$

where $\delta T = T/M$ is the time step. This expression becomes exact as $M \to \infty$ i.e. $\delta T \to 0$. In practice one uses a finite time step, which is sufficiently small to well represent the fastest varying force in the system. Regardless of the time step, the family of Trotter factorized integrators retains important symplectic and time-reversible properties.[3,35]

The above Trotter factorized integrator is identical to the traditional Velocity Verlet scheme.[36] The operator $e^{iL_p \delta T}$ can be shown to perform the operations of a momentum shift through a time interval $\delta T$:[3]

$$\{\mathbf{x}(t), \mathbf{p}(t)\} \rightarrow \{\mathbf{x}(t), \mathbf{p}(t) + \delta T\, f(\mathbf{x}(t))\} \tag{5}$$

while $e^{iL_x \delta T}$ performs a coordinate shift:

$$\{\mathbf{x}(t), \mathbf{p}(t)\} \rightarrow \{\mathbf{x}(t) + \delta T\, \mathbf{p}(t)/\mathbf{m},\ \mathbf{p}(t)\} \tag{6}$$

where **m** is the vector of masses for the degrees of freedom.

The strength of the Trotter factorization, however, is that it allows much more general decompositions of the Liouville operator. We now consider the case where the total force on each degree of freedom can be separated into fast and slow components.

$$f_i = f_i^F + f_i^S$$

Splitting the momentum shift operator, the Liouville operator can now be decomposed into three terms.

$$iL = iL_x + iL_p^F + iL_p^S = \sum_k^n \dot{x}_k \frac{\partial}{\partial x_k} + \sum_k^n f_k^F \frac{\partial}{\partial p_k} + \sum_k^n f_k^S \frac{\partial}{\partial p_k} \tag{7}$$

Proceeding as above with Suzuki-Trotter expansion, one obtains the following MTS integrator.[3]

$$\Gamma(t+\Delta T) = e^{iL_p^S \Delta T/2} \left( e^{iL_p^F \delta T/2} e^{iL_x \delta T} e^{iL_p^F \delta T/2} \right)^N e^{iL_p^S \Delta T/2} \Gamma(t) \tag{8}$$

Here, $\Delta T = N\delta T$ is the outer time step, and the bracketed term evolves the positions and momenta of the system by the smaller inner time step, $\delta T$, under the action of only the fast varying forces. The advantage of the MTS scheme is that the outer time step can be chosen with respect to the fastest *slow* force and evaluated 1/Nth as often compared to traditional Verlet integrators. Since force evaluation is the dominant computational step in AIMD calculations, the MTS approach can in principle provide an N-fold speedup over traditional integrators, as long as the appropriate decomposition in slow and fast forces can be found.

In BOMD both long- and short-range atomic forces depend on nonlinear SCF equations. Thus, a strict algebraic force decomposition of the type commonly used in classical MTS integrators is not possible. Fortunately, the RESPA strategy is flexible enough to allow a broad range of numerical decompositions. We introduce the following, almost trivial, scheme.

$$iL = \sum_k^n \dot{x}_k \frac{\partial}{\partial x_k} + \sum_k^n F_k^{mod} \frac{\partial}{\partial p_k} + \sum_k^n \left(F_k^{AI} - F_k^{mod}\right) \frac{\partial}{\partial p_k} = iL_x + iL_p^F + iL_p^S \qquad (9)$$

Here $F_k^{AI}$ is the full *ab initio* force acting on the $k^{th}$ degree of freedom while $F_k^{mod}$ is an approximate model force intended to capture the short-range behavior of $F_k^{AI}$. Assuming the model force is smooth and numerically well behaved, the resulting *ab initio* MTS integrators evolve dynamics on the same potential energy surface as traditional BOMD approaches. Of course, a poor choice of the model force could leave short-range, high-frequency components in the difference force, $F_k^{AI} - F_k^{mod}$, limiting the outer time step and defeating any speedup. However, the same logic also works in reverse. We are free to relax an exact model for the short-range force until short-range discrepancies appear in $F_k^{AI} - F_k^{mod}$ with the same timescale as the fastest long-range forces. In other words, the outer time step retains some ability to correct errors in our short-range model force. In practice we will show that sufficiently accurate model forces are easily computed, so it is the latter line of reasoning that is most relevant.

Our first approach to model the short-range *ab initio* force is to split an extended system into small independent fragments. A very general approach would require a method to fragment macro-molecules across covalent bonds as well as automatically distribute electrons among fragments. For simplicity, the present work focuses on systems where these refinements are not needed, in particular, large water clusters. In the method we christen MTS-FRAG, the model force is constructed as a sum of independent *ab initio* gradient calculations on each water molecule in vacuo. Because all molecular calculations require equal computational effort, the work to compute the model force scales linearly with the size of the system (i.e., number of fragments). On the other hand, the global *ab initio* gradient needed in the outer time step requires computational effort that is at least quadratic in the system size [for conventional implementations of Hartree-Fock (HF) or density functional theory (DFT)]. Thus, for large systems, the short-range force calculation becomes essentially free and we expect the overall speedup to increase linearly with the size of the outer MTS time step. The same analysis applies to other *ab initio* methods such as perturbation theory and coupled cluster. In these cases, the scaling of effort with respect to molecular size is often considerably steeper, which will make our MTS-FRAG scheme even more efficient compared to an implementation where the *ab initio* force is evaluated using HF or DFT.

As presented above, the fragment approach completely ignores all intermolecular interactions during the inner time steps. It is illustrative to also consider a simple refinement, in which we add Lennard-Jones and Coulomb terms between all $W$ water molecules to the model force.

$$F_k^{LJ} = \nabla_k \sum_{m=1}^{W} \sum_{n=m+1}^{W} \left( \frac{A}{r_{O_m O_n}^{12}} - \frac{C}{r_{O_m O_n}^{6}} + \sum_{i \in m} \sum_{j \in n} \frac{q_i q_j}{r_{ij}} \right) \qquad (10)$$

We denote an MTS method using this Lennard-Jones augmented fragment model force as MTS-LJFRAG. In the present work, we used empirical parameters ($A$, $C$, and charges $q_k$) directly from the TIP3P water model without modification.[37] Evaluation of this term adds negligible effort to the more expensive *ab initio* fragment calculations but allows atoms to avoid strongly repulsive regions near neighboring molecules during inner integration steps. Polarization and charge transfer effects are still neglected, but these are expected to vary on longer time scales. Further refinement of these fragment models is no doubt possible. However, we introduce these coarse approximations here in order to explore the ability of the outer time step to correct for small but significant errors in the short-range model force.

The primary limitation of the fragment approaches described above is that the atomic decomposition into fragments cannot be adjusted during the course of a simulation without destroying the reversibility and energy conservation of the integrator. This rules out arbitrary bond rearrangements, one of the key features we would like to preserve from traditional AIMD. As an alternative we consider the MTS-CASE scheme, where the model force is obtained from an electronic structure calculation with a truncated (i.e. short-range) Coulomb operator. This is accomplished with the following substitution in the electronic Hamiltonian (and also in the Coulomb interaction between the nuclear point charges).

$$\frac{1}{r} \rightarrow \frac{\text{erfc}(\omega r)}{r} \qquad (11)$$

Here $\omega$ is a constant parameter with units of inverse distance that determines the range at which the Coulomb interaction is effectively quenched. Because the CASE model force does not require an explicit molecular decomposition, it should, in principle, have no difficulty with bond rearrangements. However, the accuracy with which the CASE approximation can describe distorted transition geometries is a serious concern. Fig. 1 shows potential energy scans at the restricted Hartree-Fock (RHF) and CASE levels of theory for the dissociation of a hydroxide ion and water molecule. While CASE accurately reproduces the RHF minimum-energy oxygen-oxygen distance, the binding energy is catastrophically underestimated. Such artifacts imply that physically relevant trajectories cannot be obtained from the CASE potential surface. However, in MTS-CASE the outer time step corrects for the difference between the CASE potential and the full RHF surface. Hence the MTS-CASE trajectory evolves on the full potential surface and so should accurately describe transition geometries.

**Results and Discussion**

In order to test these approaches, we implemented the *ab initio* MTS methods described above in a development version of the TeraChem code for quantum chemistry and *ab initio* molecular dynamics.[38] We first simulated the dynamics of an $(H_2O)_{57}$ cluster at the RHF/3-21g level of theory. In all calculations, the cluster was confined to a density of 1g/mL by applying a spherical quadratic repulsive potential to oxygen atoms beyond the sphere's radius. The barrier potential was included in the inner time step of the MTS trajectories due to its negligible computational cost compared to the *ab initio* force evaluation. The system was first equilibrated for 5 ps using a standard Velocity Verlet[36] integrator with the Bussi-Parinello thermostat.[39] During equilibration we used an MD time step of 1.0 fs, the target temperature was 350K, and the thermostat time constant was 100 fs.

After equilibration, we collected a series of 21ps micro-canonical (NVE) trajectories using a range of time steps for each of the MTS-FRAG, MTS-LJFRAG, and MTS-CASE integrators outlined above. All simulations were started from identical initial conditions. Baseline dynamics using the standard Velocity Verlet integrator were collected with time steps between 0.5 and 1.5fs, since higher values lead to unstable trajectories. *Ab initio* MTS trajectories used an inner time step of 0.5fs, with outer time steps ranging from 1.5 to 3.0fs. For the MTS-CASE integrator, two screening ranges were tested: $\omega=0.33$ Bohr$^{-1}$, which corresponds to an effective screening distance of only 1.6 Å, and $\omega=0.18$ Bohr$^{-1}$, which corresponds to 5.5Å .

The drift in total energy over the course of a simulation provides a test for the quality of an integrator. An energy curve for an MTS-LJFRAG simulation using an outer time step of 2.5fs is shown in Fig. 2. The drift was extracted by performing a linear fit of the total energy over the 21ps trajectory and is almost unnoticeable on the scale of the natural kinetic energy fluctuations of the system. The extracted slope is $2.74 \times 10^{-5}$ kcal/mol/ps/dof, i.e. during the entire 21ps trajectory $5.75 \times 10^{-4}$ kcal/mol of energy is added to each degree of freedom due to inaccuracies in the integration of the equations of motion. To put this in perspective, this is equivalent to a 0.289 K rise in temperature of the system over that time. Current AIMD trajectories are generally limited to around 100ps and hence even on this time-scale the increase in temperature due to the integrator would be just over 1K. This is likely to cause a negligible difference in any desired properties and could be removed by an extremely gentle thermostat.

Fig. 3 compares the energy conservation of each MTS trajectory. All MTS methods perform well up to an outer time step of 2.5fs, which is more than double the maximum stable time step possible with a standard Velocity Verlet integrator. MTS-CASE with $\omega = 0.33$ screening is approaching the edge of acceptability. This is not surprising considering the relatively short-range interactions that the difference potential describes when the model force is screened so severely. Increasing the screening distance (decreasing $\omega$) allows MTS-CASE to be tuned to an acceptable level of energy conservation. The success of the MTS integrators suggests that the inner model forces accurately include the high frequency components of the interactions, e.g. stretches and bends, leaving a smooth slowly varying force at the outer time step. When the outer time step is increased to 3fs, the performance of all the MTS methods degrades significantly.

The failure of all methods with a 3fs outer time step is not surprising since MTS integrators such as RESPA are known to suffer from non-linear resonance instabilities.[40,41] These instabilities, which arise due to interactions between the fast forces and long time steps, mean that the fastest forces in the system limit how infrequently the slowest interactions must be calculated at the outer time step. In the case of our system the fastest modes, due to OH stretches of the dangling hydrogen atoms at the surface of the cluster, oscillate at 4000 cm$^{-1}$ (Fig. 4), corresponding to a time period of $\tau$=8.5 fs. For a harmonic oscillator one can show that the maximum stable outer time step is given by $\Delta T_{max} = \tau/\pi$ which yields a resonance limit on the time step of 2.7fs.[42,43] This matches our observations precisely, and suggests that the present limitations of our method do not stem from inadequacy of the model potentials. It has been shown that resonance instabilities can be effectively mitigated using specially designed Langevin thermostats.[41,44] The application of these techniques here is beyond the scope of the

Luehr, Markland and Martínez – Page 12

present study, but should increase the outer time step by a further factor of 2-4 fold and thus further improve the computational speedups reported here.

While energy conservation provides a test of the stability of MTS methods one would also like to ensure that more subtle properties of the system remain unchanged. For example, Fig. 4 shows the power spectrum of the system using 0.5fs and 1.0fs time steps with a traditional Velocity Verlet integrator. Although in Fig. 3 we found these both to provide acceptable energy conservation, the 1.0fs Verlet integrator shows a clear frequency shift at high frequencies compared to the 0.5fs Verlet and MTS integrators. Hence, although using standard Velocity Verlet may be stable, one should take care regarding the properties obtained when large time-steps are employed.

Figure 5 shows the power spectra from the MTS trajectories using a 2.5fs outer time step and 0.5fs inner time step compared with that obtained using standard Velocity Verlet with a 0.5fs time step. In all cases the agreement is very good, with the peak positions being very well reproduced. Remarkably all MTS methods with a 2.5fs outer time step capture the power spectrum better than a Verlet integrator using a 1.0fs time step. Peak intensities show greater variation, though this is likely due to the large statistical error bars on power spectra obtained from a single 21ps trajectory.

It is clear from Fig. 5 that the empirically fit terms in the LJFRAG model improve on the simpler FRAG approach. This suggests that the unmodified TIP3P force field is capturing some of the relevant high frequency intermolecular forces, such as those from hydrogen bonds that are not included by the simpler monomer fragment approach. Among the CASE spectra, the $\omega=0.18$ model shows only marginal improvement over the coarser ω=0.33 cutoff. Both outperform the simple fragment approach, and are roughly equivalent to the Lennard-Jones fragment model.

It is also instructive to compare MTS-CASE dynamics with that obtained by traditional (Velocity Verlet) integration on the CASE potential surface. Fig. 6 compares the power spectrum obtained from a 14ps simulation of dynamics on the CASE (ω=0.33) potential energy surface using a 0.5fs Velocity Verlet integrator to that obtained from 0.5fs (RHF) Verlet and 2.5fs/0.5fs MTS-CASE (ω=0.33). The Verlet-CASE power spectrum is in pronounced disagreement with that obtained from dynamics on the RHF surface. This is not surprising given the major quantitative differences between the RHF and CASE potential surfaces shown in Fig. 1. It is notable that the differences are most pronounced at the highest frequencies since one might expect the O-H stretch to be relatively well preserved within CASE, given that the Coulomb operator is unmodified at short ranges. However, this peak shows much less structure in the CASE spectra and is significantly red-shifted from 4000 to 3600cm$^{-1}$. Given these significant differences, it is remarkable that the MTS-CASE trajectory gives a power spectrum close to that obtained on the RHF surface despite using the CASE potential as the model for high-frequency force updates.

Thus far we have only considered water clusters, which do not undergo covalent bond rearrangements on the simulated time-scale. However, one of the main advantages of AIMD simulation is the ability for the system to undergo spontaneous covalent bond breaking and formation during the trajectory. Hence to evaluate the ability of our MTS-CASE approach to describe bond rearrangements, we simulated the proton transfer dynamics of a hydroxide ion solved in a cluster of 64 water molecules. The system was equilibrated using the same procedure and boundary conditions as the $(H_2O)_{57}$ cluster described above. A 21ps microcanonical simulation was run at the RHF/3-21g level of theory using the MTS-CASE($\omega$=0.33) integrator. Fig. 7 shows a 2ps window of the total trajectory. The energy drift of $1.54 \times 10^{-4}$ kcal/mol/ps/dof (0.077K/ps/dof) was calculated by a linear fit to the entire 21ps trajectory. For each frame the hydroxide oxygen atom was determined by assigning each hydrogen atom to the nearest oxygen atom and then reporting the index of the oxygen atom with a single assigned hydrogen (red line in Fig. 7). As shown in the upper panel of Fig. 7, a proton oscillates between primarily two oxygen atoms (shown in blue in the lower panel of Fig. 7). During the entire 21ps trajectory we observed over 500 proton transfer events. However, the total energy drift is comparable to that reported above for the nonreactive $(H_2O)_{57}$ cluster. This is remarkable on two counts. First, it demonstrates that the MTS scheme is applicable even when bonds are being formed and broken – a difficult task for empirical force fields and thus also for the fragment-based MTS-FRAG and MTS-LJFRAG schemes described above. Secondly, the CASE approach we are using here is the most aggressively screened of the two we have considered. With a screening factor of $\omega$=0.33 bohr$^{-1}$, the Coulomb interaction between any two charged particles is already being attenuated at a distance of 1.6Å. Hence, in the inner time step, the proton barely interacts with any electronic orbitals other than those on the two nearest heavy atoms. All other interactions are corrected in the outer time step, which demonstrates remarkable robustness by its ability to maintain the energy conservation observed above.

Finally, we consider the computational efficiency of our *ab initio* MTS approach based on our initial implementation in TeraChem. Table 1 summarizes the performance of the Lennard-Jones fragment MTS scheme relative to a standard Velocity Verlet integrator with a time step of 0.5fs. Even though the Velocity Verlet integrator is stable with a 1fs time step (Fig. 3), such a large time step noticeably alters the dynamics, as demonstrated by the shift of the high frequency peak in the power spectrum (Fig. 4). Hence, Velocity Verlet with a 0.5fs time step is the appropriate comparison. These calculations were carried out with a larger $(H_2O)_{120}$ cluster treated at the RHF/6-31g level of theory and confined to a density of 1g/mL by a spherical boundary. All calculations used a single Tesla C2070 GPU. Because the outer time step still requires a full SCF gradient evaluation, the best that can be achieved is a 5x speedup, if we compare an MTS integrator with a 2.5fs outer time step to a standard Velocity Verlet method with a time step of 0.5fs. We are able to achieve up to 4.4x speedup, which is over 88% efficient.

As with the fragment models, the CASE approximation leads to linear scaling computational effort in evaluating the inner time steps.[32] This is now due to improved screening of electron repulsion integrals afforded by a short-range coulomb operator. The present experimental code does not yet exploit the improved CASE screening. However, previous work has shown that CASE at the screening levels employed here offers significant computational speed-ups.[32]

**Conclusion**

In this paper we have demonstrated new approaches that allow MTS integrators to be applied generally to AIMD calculations. We exploited the ability of the outer MTS time steps to correct low frequency modeling errors within the inner time steps. Thus we were able to employ drastically simplified short-range approximations to the *ab initio* forces in the inner time steps. Despite these computational savings, the resulting methods remain robust, exhibiting symplecticity and time-reversibility and providing excellent energy conservation and even improved dynamical properties compared to Verlet integrators with moderately large time steps. As with single-step integrators our *ab initio* MTS methods can be systematically improved by reducing the time steps or, for MTS-CASE, also by increasing the screening distance.

The model forces used here are inspired by linear scaling approximations. However, while linear scaling methods attempt to globally capture all significant interactions, MTS model forces need only represent the highest frequencies within the system, and even here low frequency differences between the model and *ab initio* systems can be tolerated. This means that much looser thresholds can be employed in the case of MTS model forces. Thus, even for systems too small to reach the crossover point for traditional linear scaling methods, our MTS scheme can provide a significant speedup. For larger systems where linear scaling approaches are applicable, the MTS approaches introduced here can still be applied, since its looser thresholds should allow the model forces to be computed more cheaply than the more accurate linear-scaling gradient.

Other model forces are obviously possible. For example, our use of the TIP3P force field suggests a completely empirical model force. However, the benefits of such models are limited by the cost of the global gradient calculation in the outer time step. In the present case, reducing the computational effort of the inner time step would at best yield a 13% speedup over the fragment approaches described above. Thus, future work should focus on extending the outer time step and on reducing the cost of the global gradient evaluation. For example, employing existing Langevin thermostats[44] to remove spurious resonance effects should allow the outer time step to be extended to 10 or even 20 fs and provide even more substantial performance gains.

**Acknowledgments**

This work has been supported by the National Science Foundation (OCI-1047577) and by the Department of Defense (Office of the Assistant Secretary of Defense for Research and Engineering) through a National Security Science and Engineering Faculty Fellowship (NSSEFF) to TJM. T.E.M acknowledges funding from a Terman fellowship and Stanford start-up funds.

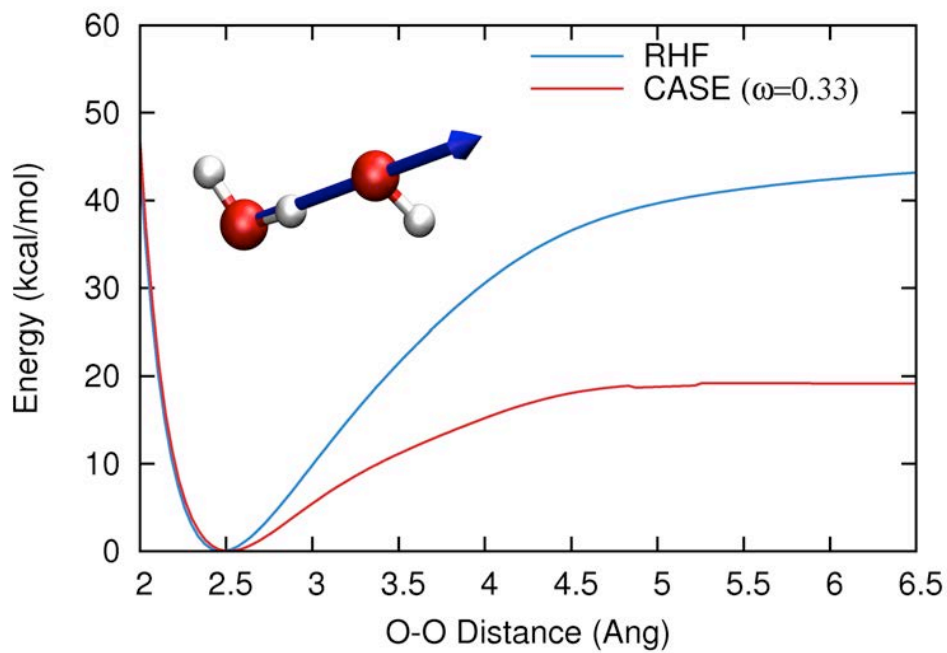

**Figure 1:** Dissociation curves for a $H_2O/OH^-$ cluster at the CASE and RHF levels of theory using the 6-31g basis set. Potential curves are generated from optimized geometries at constrained oxygen-oxygen distances. The energy for each curve is taken relative to its minimum value. The equilibrium bond distances are very similar. However, CASE severely under-estimates the binding energy, and includes an unphysical kink near 5 Anstroms.

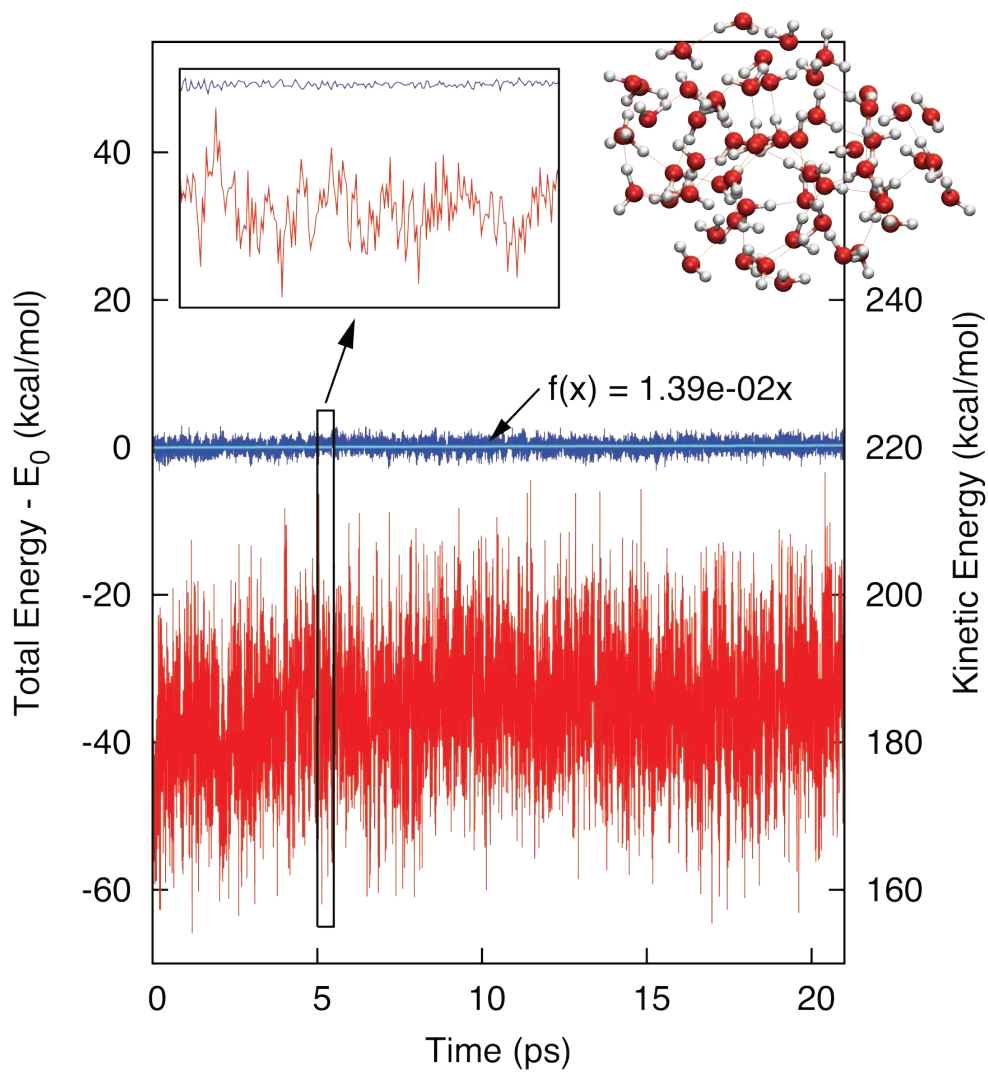

**Figure 2:** Energy conservation for 21ps simulation of an $(H_2O)_{57}$ water cluster using the MTS-LJFRAG integrator with outer and inner time steps of 2.5 and 0.5fs respectively. The simulation was run at the RHF/3-21g level of theory in the NVE ensemble after 5ps NVT equilibration at 350K. The cluster was confined by a spherical boundary chosen to lead to a density of 1g/mL. The blue curve shows total energy in kcal/mol shifted by $+2.72 \times 10^6$ kcal/mol. The cyan line is a least-squares fit of total energy to show drift. Slope is $1.39 \times 10^{-2}$ kcal/mol/ps for all degrees of freedom, i.e. $2.74 \times 10^{-5}$ kcal/mol/ps/dof. The red curve shows total kinetic energy for scale.

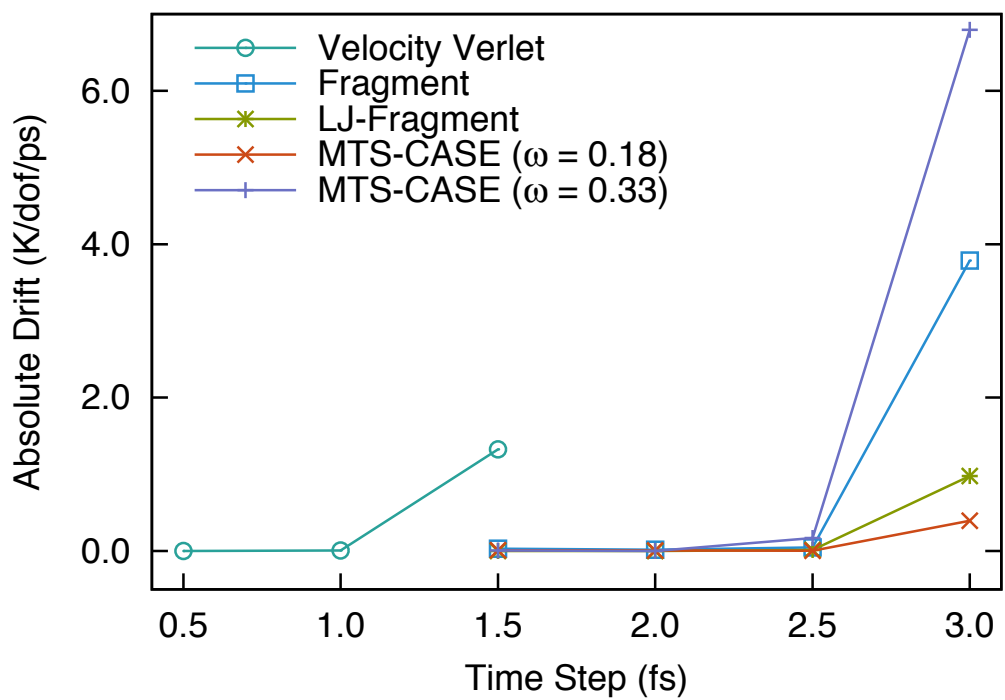

**Figure 3:** Energy drift in units of Kelvin per degree of freedom per ps for a range of integration time steps. Drifts calculated as slope of least squares fit to total energy from a 21ps NVE trajectory simulation of $(H_2O)_{57}$ at the RHF/3-21g level of theory. All trajectories were started from the same initial conditions, generated by 5ps NVT simulation at 350K. MTS trajectories used an inner time step of 0.5fs.

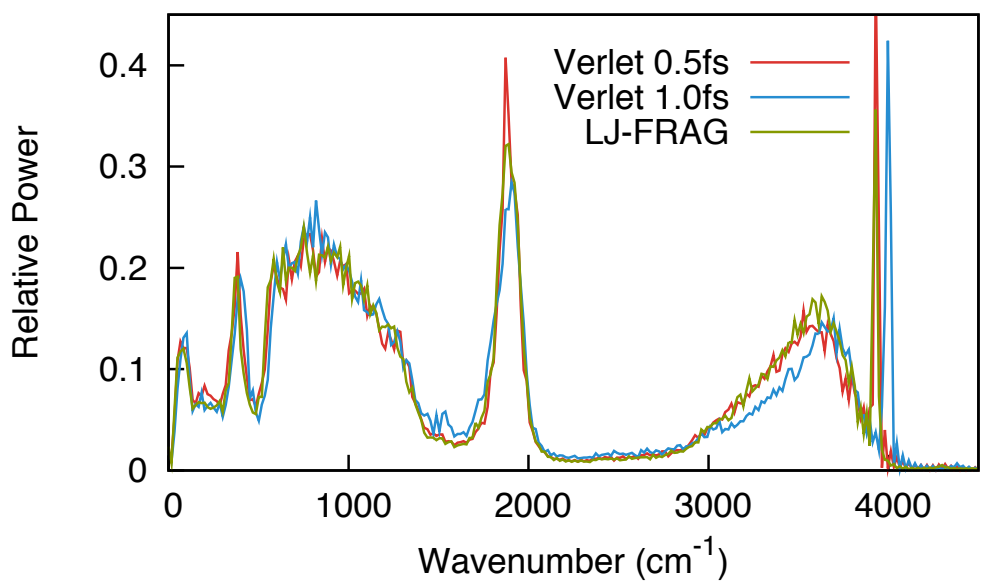

**Figure 4:** Power spectrum comparison between Velocity Verlet with 0.5fs time step (red), Velocity Verlet with 1.0fs time step (blue) and MTS-LJFRAG integrator with 2.5 and 0.5fs outer and inner time steps respectively (green). Spectra are based on 21ps NVE trajectories at the RHF/3-21g level of theory. System consists of 57 water molecules confined by a spherical boundary to a density of 1g/mL.

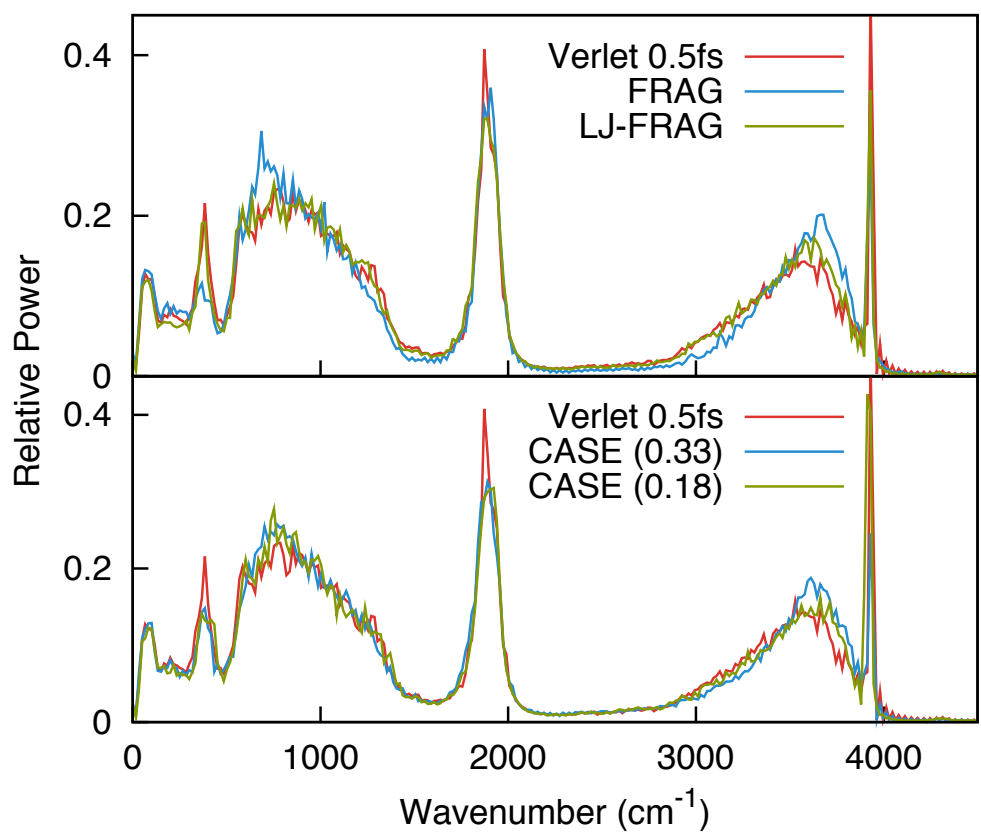

**Figure 5:** Power spectrum comparison between standard Velocity Verlet (red), and MTS trajectories. Upper box compares simple fragment (blue), and Lennard-Jones augmented fragment (green) methods. Lower box compares MTS-CASE integration using two values of omega, 0.33 (blue) and 0.18 (green), which represent 1.6 and 3Å cutoffs of the Coulomb potential respectively. Spectra are based on 21ps NVE trajectories at the rhf/3-21g level of theory. System consists of 57 water molecules confined by a spherical boundary to a density of 1g/mL. MTS integrators use 2.5fs and 0.5fs for outer and inner time steps respectively. Verlet integrator uses 0.5fs time step.

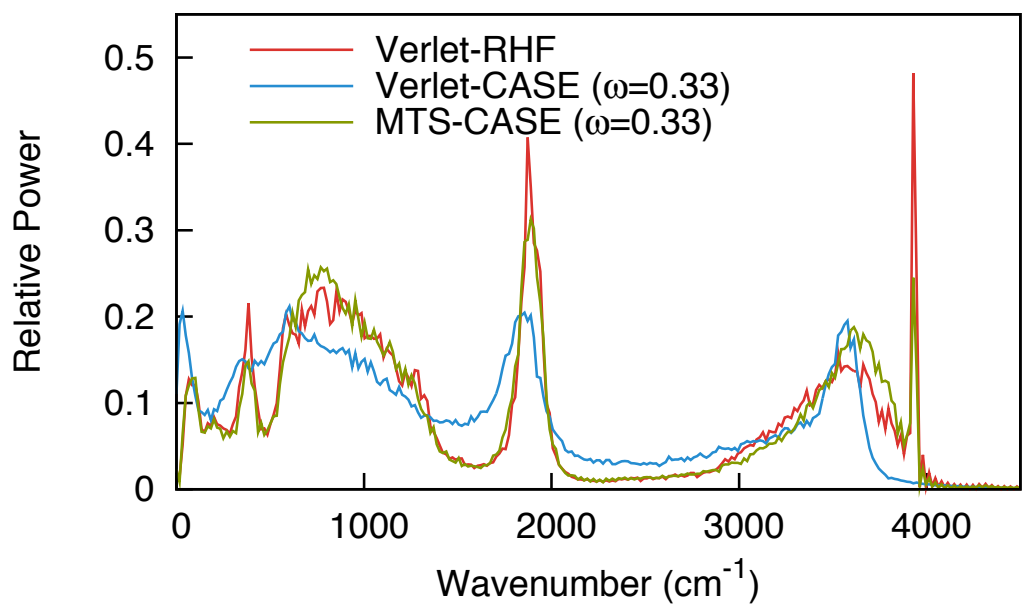

**Figure 6:** Power spectra resulting from Velocity Verlet integrated dynamics on the CASE (blue) and RHF (red) potential energy surfaces compared with MTS-CASE (green) integration. The CASE approximation uses a cutoff of 0.33 Bohr$^{-1}$ (1.6 Å) for both the Verlet and MTS integrators. Verlet-RHF and MTS-CASE spectra are based on 21ps NVE trajectories using the 3-21g basis set. Verlet-CASE spectrum is based on a shorter 14ps trajectory at the same level of theory. The system consists of 57 water molecules confined by a spherical boundary to a density of 1g/mL. Outer and inner timesteps for the MTS integrator are 2.5fs and 0.5fs respectively. Verlet integrators use 0.5fs time steps.

Luehr, Markland and Martínez – Page 24

| Integrator | Time Step (fs) | Time/Step (sec) | Steps/Day | fs/day | Speedup |
| --- | --- | --- | --- | --- | --- |
| Velocity Verlet | 0.5 | 430 | 201 | 100 | 1.0 |
| Velocity Verlet | 1.0 | 422 | 204 | 204 | 2.0 |
| MTS-LJFrag | 1.5 | 460 | 188 | 282 | 2.8 |
| MTS-LJFrag | 2.0 | 471 | 183 | 367 | 3.7 |
| MTS-LJFrag | 2.5 | 485 | 178 | 446 | 4.4 |

**Table 1:** Performance of MTS Lennard-Jones fragment compared to standard Velocity Verlet integrator. The system consists of 120 water molecules confined by a spherical boundary to a density of 1g/mL. The simulation was run at the RHF/6-31g level of theory using a single NVIDIA Tesla C2070 GPU. Step sizes for MTS integrator are for outer time step; 0.5fs inner time steps were used throughout. Timings are averaged over 100 MD steps and are in units of wall-time seconds per outer step. Speedups are computed by comparison to the 0.5fs Velocity Verlet integrator.

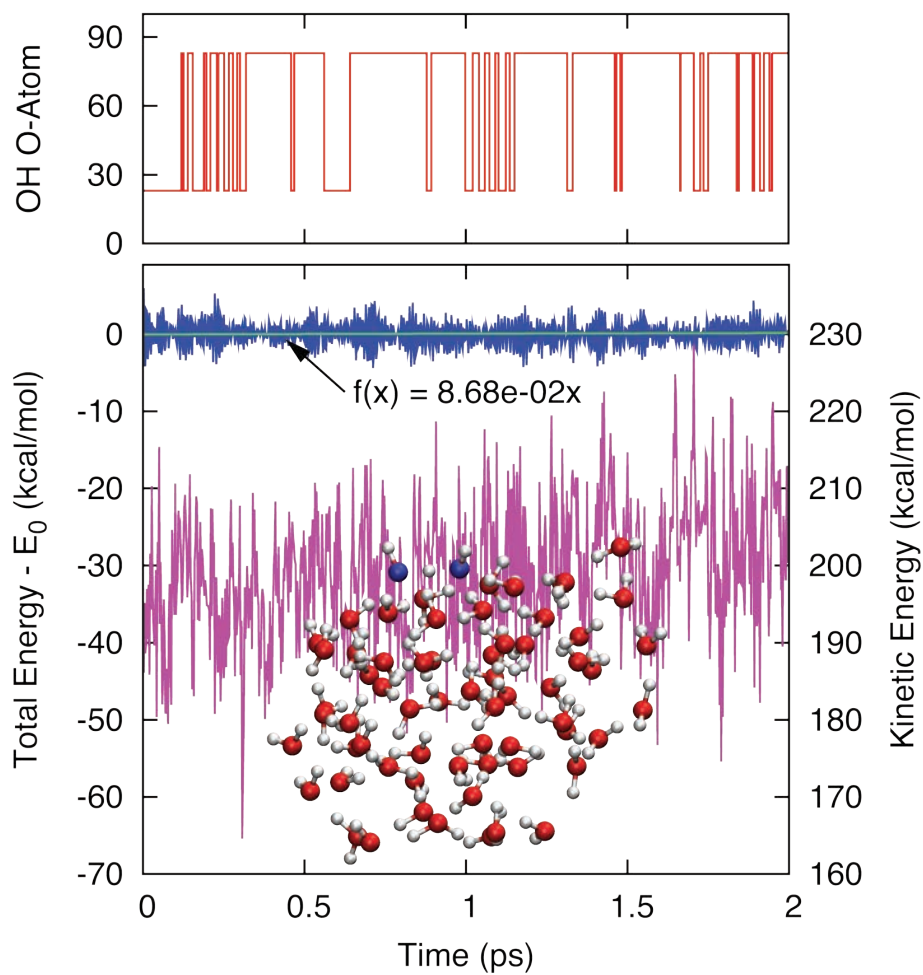

**Figure 7:** Energy conservation of MTS-CASE integrator. Plot shows 2ps window from a longer 21ps trajectory. The NVE simulation was run at the RHF/3-21g level of theory after 5ps of NVT calibration to 350K. The inner and outer time steps were 0.5 and 2.5fs. The Coulomb attenuation parameter was 0.33 Bohr$^{-1}$. The system was made up of a hydroxide ion solvated by 64 water molecules and was confined by a spherical barrier to a density of 1g/mL. A representative snapshot shows a proton transition between two oxygen atoms highlighted in blue. The blue curve shows the total energy in kcal/mol shifted by -3.10x10$^6$ kcal/mol. The green line gives a least squares fit of the total energy curve for the entire 21ps trajectory. Its slope records a drift of 8.7x10$^{-2}$ kcal/mol/ps for the entire system or 1.5x10$^{-4}$ kcal/mol/ps/dof. The red line indicates the index of the hydroxide oxygen atom for each time step. This was determined by assigning each hydrogen atom to its nearest oxygen neighbor, and then reporting the oxygen with a single assigned hydrogen. The magenta curve shows the total kinetic energy for scale.